\documentstyle[12pt]{article}

\fussy
\def\be{\begin{equation}}

\def\ee{\end{equation}}

\def\bea{\begin{eqnarray}}

\def\eea{\end{eqnarray}}

\newcommand{\ba}{\begin{array}}
\newcommand{\ea}{\end{array}}

\def\lsim{\mathrel{\rlap{\lower4pt\hbox{\hskip1pt$\sim$}}
    \raise1pt\hbox{$<$}}}	  
\def\gsim{\mathrel{\rlap{\lower4pt\hbox{\hskip1pt$\sim$}}
    \raise1pt\hbox{$>$}}}	  

\def\Pom{{\bf I\!P}}

\def\lsim{\mathrel{\rlap{\lower4pt\hbox{\hskip1pt$\sim$}}
    \raise1pt\hbox{$<$}}}         
\def\gsim{\mathrel{\rlap{\lower4pt\hbox{\hskip1pt$\sim$}}
    \raise1pt\hbox{$>$}}}         

\def\Pom{{\bf I\!P}}

\begin{document}





\begin{center}
DIFFRACTIVE DIS FROM THE COLOR DIPOLE BFKL POMERON
\footnote{To be published in the Proceedings of the DIS96
Workshop, 19-13 April 1996, Roma}
\bigskip\\

{ N.N.NIKOLAEV$^{a),b),c)}$, B.G.ZAKHAROV$^{c)}$ }
\\

{$^{a)}$ITKP der Universit\"at Bonn, Nusallee 14-16, D-53115 Bonn\\
$^{b)}$IKP, KFA J\"ulich, D-52425 J\"ulich, Germany\\
$^{c)}$L.D.Landau Institute, Kosygina 2, 1117 334 Moscow, Russia}

\end{center}




{
We review the recent progress in the theory
of diffractive DIS, focusing on predictions of strong breaking
of the Ingelman-Schlein-Regge factorization and the related
breaking of the GLDAP evolution for the diffractive structure
function. }
\section{Diffractive DIS and partonic structure of the photon}
The microscopic QCD mechanism of diffractive DIS (DDIS) is a grazing,
quasielastic scattering of multiparton Fock states of the photon
on the target proton, which is best described
viewing these Fock states as systems
of color dipoles spanned between quarks, antiquarks
and gluons \cite{NZ91,NZ92,NZ94}.
In inclusive DIS, the effect of gluons in the photon is reabsorbed
into the
Bjorken $x$ dependence of the color dipole
cross section $\sigma(x,r)$, which satisfies the running gBFKL equation
\cite{NZ94,gBFKL}. In the pQCD domain
of small dipoles $r$
\be
\sigma(x,r)={\pi^{2}\over 3}r^{2}\alpha_{S}(r)G(x,q^{2}
\approx {10\over r^{2}})
\label{eq:1.1}
\ee
where $G(x,q^{2})$ is the gluon structure function of the
target nucleon \cite{NZglue}.

A color dipole treatment of DIS comes truly of age in DDIS.
The driving term of DDIS is
an excitation of the $q\bar{q}$ state for which
\cite {NZ92} (we focus on $t=0$)
\vspace{-.1cm}\\
\be
{ d \sigma^{D} \over dt } =
\int dM^{2}{ d \sigma^{D} \over dt dM^{2} } =
 {1\over 16 \pi}\int_{0}^{1}dz \int d^{2}\vec{\rho}\,
\vert\Psi_{\gamma^{*}}(Q^{2},z,r)\vert^{2} \sigma^{2}(x_{\Pom},r)\, ,
\label{eq:1.2}
\ee
\vspace{-.1cm}\\
where $|\Psi_{\gamma^{*}}|^{2}$ is the color dipole distribution
in the photon \cite{NZ91}.
Eq.~(\ref{eq:1.2}) and its generalizations to higher Fock states are
rigorous field theory results and
answer all the  pertinent questions
\cite{NZ94,NZsplit,GNZ95,GNZcharm,GNZlong,GNZA3Pom}
:
i) how DDIS scales with $Q^{2}$?
ii) does the partonic
structure function of the pomeron make any sense?
iii) is DDIS soft or hard process? iv) how to
control the hardness of DDIS?
v) what is the flavor content of DDIS?
vi) what is $\sigma_{L}/\sigma_{T}$ for DDIS?
vii) how inclusive DDIS matches exclusive vector meson
production? In our understanding of DDIS we are  entering the era
of enlightenment, one must focus on adequate Monte
Carlo implementation of QCD ideas on DDIS (to be reported at
the next
DIS workshop?), see Ada Solano's
status report at this conference \cite{Ada}.

\section{Poor man's interpretation of diffraction: the
Regge factorization
}
The diffractive structure function is operationally defined as
\vspace{-.1cm}
\be
(M^{2}+Q^{2})
{ d\sigma^{D}
\over dt\,d M^{2} } =
{  \sigma_{tot}(pp) \over 16\pi}
{4\pi^{2} \alpha_{em}
\over Q^{2}}\left\{
 F_{T}^{D}(x_{\Pom},\beta,Q^{2})
+\varepsilon_{L} F_{L}^{D}(x_{\Pom},\beta,Q^{2})\right\}\, .
\label{eq:2.1}
\ee
The variables $\beta =Q^{2}/(Q^{2}+M^{2})$ and
$x_{\Pom}=(Q^{2}+M^{2})/(Q^{2}+W^{2})=x/\beta$
(we stick to the Eilat convention, $\varepsilon_{L}$ is the longitudinal
polarization of the photon) don't tell
anything of the interaction mechanism and
the words "pomeron exchange" bear no information beyond
that the reaction considered is diffractive.
Still, the DIS community fell the temptation of endowing
the pomeron with
the usual attributes of a particle such as the partonic structure
functions $F^{\Pom}_{T,L}(\beta,Q^{2})$ and the flux of pomerons
$\phi_{\Pom}(x_{\Pom})/x_{\Pom}$ in the proton \cite{Ingelman}:
\vspace{-.2cm}
\be
F_{T,L}^{D}(x_{\Pom},\beta,Q^{2})=
\phi_{\Pom}(x_{\Pom})
 F_{T,L}^{\Pom}(\beta,Q^{2})\, .
\label{eq:2.2}
\ee

\section{Diffraction defies the Ingelman-Schlein-Regge
factorization, is mostly soft  but exhibits the
Bjorken scaling
}

The Ingelman-Schlein-Regge factorization (\ref{eq:2.2})
is not borne out
by the QCD theory of diffraction, which we always maintained
since our 1991 work \cite{NZ92} and demonstrated explicitly
in 1994 \cite{NZsplit}.
First, focus on $q\bar{q}$ excitation.  For
T photons one finds $
{ d \sigma_{T}^{D}/ dt } \propto
 G^{2}(x_{\Pom},q_{T}^{2}\approx m_{f}^{2})/Q ^{2}m_{f}^{2}$,
which is dominated by the contribution from soft dipoles
$r\sim 1/m_{f}$. Beware of the nonperturbative
contributions on top of the perturbative gBFKL pomeron,
still the  $1/Q^{2}$ leading twist behavior of
$\sigma_{T}^{D}$ is a rigorous result \cite{NZ91,NZ92},
neither the Regge factorization nor the concept of the
partonic structure of the pomeron are needed for that!
A comparison with
(\ref{eq:2.2}) implies a plethora of flavor
dependent "pomeron flux factors" \cite{GNZ95,GNZcharm}
$
\phi_{val}(x_{\Pom}) \propto G^{2}(x_{\Pom},
q_{T}^{2}\approx m_{f}^{2})\, ,
$
in defiance of the Ingelman-Schlein-Regge factorization.
For L photons \cite{NZ92,GNZlong}
$\sigma_{L}^{D} \propto
G^{2}(x_{\Pom},q_{L}^{2}\approx {1\over 4}Q^{2})/Q^{4} $.
It is dominated by $r\sim 1/Q$ and has  the higher twist
behavior. The pQCD scale
$q_{L}^{2}\approx {1\over 4}Q^{2} \neq q_{T}^{2}$, in
further defiance of the Ingelman-Schlein-Regge factorization,
which must be buried in state.
\vspace{-.2cm}


\section{Dijets and more on the factorization breaking
}

Excitation of the $q\bar{q}$ state gives rise to the
back-to-back jets with the transverse momentum
$\vec{k}$ with respect to the $\gamma^{*}\Pom$ collision
axis. The jet cross sections have been derived in
\cite{NZ92,NZsplit} in a simple analytic form
 and elaborated in \cite{GNZcharm,GNZlong}.
One finds
$
d\sigma_{T,L}^{D}/dM^{2}dk^{2}dt \propto G^{2}(x_{\Pom},q^2_{T,L})\,
$
with the $\beta,k^{2}$ dependent pQCD scale $q_{T,L}^{2}$ which
increases towards
$M^{2}\ll Q^{2}$ and/or large $k^{2}$:
\be
q_{T,L}^{2}=(k^{2}+m_{f}^{2})(M^{2}+Q^{2})/ M^{2}
=(k^{2}+m_{f}^{2})/(1-\beta)
\label{eq:4.2}
\ee
Remarkably, $d\sigma_{L} \propto 1/k^{2}$ and
$\sigma_{L}$ is dominated by large angle jets with
$k^{2}+m_{f}^{2} \approx {1\over 4}M^{2}$.
For dijets with $M^{2} \gg Q^{2}$ and real photoproduction
see \cite{NZsplit}. Bartels reported on similar results
\cite{Bartels} which are also
based on our technique \cite{NZsplit}.\vspace{-.2cm}

\section{The valence quarks and gluons and sea of the pomeron}

$F_{T}^{D}$ for $q\bar{q}$ excitation resembles the valence
structure function of hadrons
\cite{NZ92,NZ94,GNZcharm}
\be
F_{T}^{D,val}(x_{\Pom},\beta,Q^{2}) \propto 0.27\beta(1-\beta)\, ,
\label{eq:5.1}
\ee
In a very narrow limit of $\beta \rightarrow 1$ one
finds $F_{T}^{D,val}\propto (1-\beta)^{2}$, see \cite{GNZcharm}.
With the grain of salt and due reservations for the breaking of the
Ingelman-Schlein-Regge factorization,
$q\bar{q}$ excitation
can be interpreted as DIS on the valence $q\bar{q}$
of the pomeron. For the
flavor content and normalization in (\ref{eq:5.1}) see
\cite{NZ92,GNZ95}.

The $q\bar{q}g$ excitation is a driving term of DDIS
at $\beta \ll 1$. The crucial finding
is a dominance of the leading log$Q^{2}$
ordering of sizes \cite{NZ94}
\be
Q^{-1} \lsim r \ll \rho \sim R_{c}\, ,
\label{eq:5.2}
\ee
where $r$ and $\rho$ are the $q\bar{q}$ and $qg$ separations.
Here $R_{c}\sim 0.3$fm is
the radius of propagation of perturbative gluons in the
nonperturbative QCD vacuum and defines still another
factorization scale $q^{2}\approx \mu_{G}^{2}=R_{c}^{-2}$,
which is universal for all the
flavors and for L and T photons.
The factor $\beta$ in $F_{T}^{D,val}$ derives
from the spin ${1\over 2}$ of quarks.
For the presence of a spin 1 gluon in the $q\bar{q}g$ state of
the photon, one finds  \cite{NZ92,NZ94,GNZ95}
$
F_{T}^{D,sea}(x_{\Pom},\beta,Q^{2})\sim (1-\beta)^{2}
G^{2}(x_{\Pom},
q_{sea}^{2}\approx \mu_{G}^{2})\, ,
$
which is approximately flat at $\beta \ll 1$ \cite{NZ92,NZ94}.
As such it resembles the sea structure function of hadrons, and
excitation of the $q\bar{q}g$ state can be interpreted as DIS
on the $q\bar{q}$ sea in the pomeron, which was generated from
the valence gluon-gluon component of the pomeron in precisely
the same manner as sea in hadrons evolves from gluons. For the
wave function of the $gg$ state of the pomeron
and $G_{\Pom}(\beta)\propto (1-\beta)$ see \cite{NZ94}.
Because
$\mu_{G}$ is different from the quark masses $m_{f}$, this
brings a still new specimen into the menagerie of "pomeron fluxes"
$
f_{sea}(x_{\Pom}) \propto G^{2}(x_{\Pom},q_{sea}^{2}
\approx \mu_{G}^{2})\,$, for the detailed parameterization of
different flux functions see \cite{GNZ95,GNZcharm,GNZlong}.
The normalization $g$ of the sea term in
$
F_{2}^{D}(x_{\Pom},\beta,Q^{2})\propto \beta(1-\beta)+g (1-\beta)^{2}
$
is related to the so-called triple-pomeron coupling
\cite{NZ92,NZ94,GNZA3Pom}.
The  $\propto (1-\beta)^{2}$ sea term is non-negotiable,
already in 1991 we predicted $g\sim .5$ \cite{NZ92}, for slight
update see \cite{GNZ95}, ZEUS
\cite{ZEUSF2Pom} gave an important confirmation
of our predictions for $F^{D}$ and for the sea term
in particular:
$g=0.34 \pm 0.16$. Predictions for DDIS are parameter free,
they use the same gBFKL color dipole cross section
which gives a perfect description of
the proton structure function \cite{NZHera} and of the vector
meson production \cite{NNZdipole}. There is no momentum sum rule
for the pomeron, momentum fractions
$\langle \beta_{i} \rangle$ change with $x_{\Pom}$. Our prediction
\cite{NZ92,NZ94,GNZ95} for a moderately small
$x_{\Pom}$ is that $\langle \beta_{val} \rangle \sim
\langle \beta_{sea} \rangle \sim \langle \beta_{glue} \rangle$.


\section{Chameleon exponent $n$ of the $\propto x_{\Pom}^{-n}$ fits}

The exponent $n$ of the popular fit
$F_{T,L}^{D}(x_{\Pom},\beta,Q^{2}) \propto x_{\Pom}^{1-n}$
describes the $x_{\Pom}$ dependence of the flux functions
$\phi_{val},f_{sea}$. We predicted \cite{GNZ95,GNZcharm}
that at HERA, from $x_{\Pom}
\sim 0.1$ down to $x_{\Pom} \sim 0.001$, the fluxes for the
valence light quark, valence charm and sea components of the
pomeron must diverge by the factor $\sim 2-5$ !
For instance, the abundance of charm in DDIS is
predicted to rise from $\approx 3$\% at $x_{\Pom}=0.01$ to
$\approx 25$\% at $x_{\Pom}=0.0001$. For this plethora of pomeron
fluxes, $n$ must vary with
flavor, $\beta$, $Q^{2},k^{2}$, what not, even for
the pure pomeron exchange, which is
a non-negotiable QCD prediction.


\section{Back to the triple-Regge phenomenology?}

The recent data on $n$ from ZEUS \cite{ZEUSRoma} and H1
\cite{H1Roma} are inconclusive and somewhat conflicting.
Different selections of DDIS by H1 and ZEUS may pick up
different contamination from the secondary
reggeon exchanges, in particular at a not so small
$x_{\Pom}$, which is the case at small $\beta$. For instance,
for the $\pi$-exchange
$n_{\pi}\approx - 1$ vs. $n_{\Pom} \sim 1.2-1.3$,
the $\pi$ contamination rises with $x_{\Pom}$
and is substantial already at $x_{\Pom}\sim 0.1$
\cite{HERApion}. For the $\rho,\omega,\Pom'$ exchanges
$n_{R}\approx 0$. Reggeon contributions decrease the
observed $n(\beta)$ towards small $\beta$ precisely the way
reported by H1 \cite{H1Roma}. Because the DDIS is soft
dominated (Section 3), the reggeon contamination must be
about the same as in hadronic diffraction \cite{3Regge},
i.e., non-negligible even at $x_{\Pom} \sim 0.05$.
Landshoff \cite{Landshoff} had made similar observations. The
good old triple-Regge phenomenology is called upon! We
suggest a simple test of the reggeon contamination: the
smaller is $Q^{2}$ the smaller are $x_{\Pom}$ and the
weaker must be the depletion of $n(\beta,Q^{2})$ towards
small $\beta$.

\section{$\sigma_{L}/\sigma_{T}$ for diffractive DIS}

Compare $F_{T}^{D}$ of Eq. (\ref{eq:5.1}) with
$
F_{L}^{D,val}(x_{\Pom},\beta,Q^{2})\propto {1\over Q^{2}}
(1- 2 \beta)^2 \beta ^3\, ,
$
which strongly peaks at large $\beta$ and has a specific zero at
$\beta=0.5$ \cite{NZsplit,GNZlong}. Compared to DIS on hadrons,
this is an entirely new situation.
The photon polarization $\epsilon_{L}$ can readily be varied
changing $x_{\Pom}$.
Because the $x_{\Pom}$ dependence of "pomeron fluxes" is under
good control, the predicted \cite{GNZlong}
dominance of $F_{L}^{D}$ at
$\beta \gsim 0.9$ can easily be tested experimentally.
In the sea region of $\beta \ll 1$ we predict \cite{GNZlong}
$F_{L}^{D,sea}/F_{T}^{D,sea} \approx 0.2$, which is the same
as for inclusive DIS \cite{NZHera}.


\section{The $Q^{2}$ and $\beta$ evolution of diffractive DIS
and jets at $\beta\ll 1$}

The familiar GLDAP evolution derives from the radiation
of partons with transverse momenta $R_{N}^{-2} \lsim k^{2}
\lsim Q^{2}$. In DDIS instead of fixed $R_{N}^{-2}$
there emerges the scale (\ref{eq:4.2}) which is $\sim Q^{2}$
at $\beta \rightarrow 1$. Furthermore, the interplay of the
virtual and real radiative corrections is quite different
for inclusive DIS and DDIS \cite{NZ94}, which may explain why
$F_{2}^{D}(x_{\Pom},\beta,Q^{2})$ at large $\beta$ refuses to
decrease with $Q^{2}$ \cite{H1F2Pom,ZEUSF2Pom}, more theoretical
work is needed here. The threshold effects
in the charm excitation produce \cite{GNZcharm} a non-negligible rise
of large-$\beta$ $F_{2}^{D}$ with $Q^{2}$.
At last but not the least,
$F_{L}^{D}$ dominates at large $\beta$, which is not the
case with the hadrons. The GLDAP analysis
of the $Q^{2}$ dependence of $F_{2}^{D}$
is illegitimate at large $\beta$, hasty conclusions
from such an analysis of DDIS on $G_{\Pom}(\beta)$ singular
at $\beta \rightarrow 1$ must be ignored. The sound expectations
for the $\beta$ dependence of the valence quark, glue and sea
described in Section 5 stay viable.

The situation changes profoundly at $\beta\ll 1$. Here
we have a rigorous proof \cite{NZ94} that for
the leading log$Q^{2}$ ordering of sizes (\ref{eq:4.2})
the $Q^{2}$ and $\beta$ evolution of
$F_{2}^{D}(Q^{2},x_{\Pom},\beta)$ must be similar to that
of the proton structure function, which agrees with the
experiment \cite{H1F2Pom,ZEUSF2Pom,H1Roma}.
Finally, for the same ordering of sizes (\ref{eq:4.2})
production of the quark-antiquark jets in
diffractive excitation of the $q\bar{q}g$ states
proceeds via standard fusion of photons with the valence
gluons of the pomeron and there is experimental
evidence for that \cite{ZEUSjet,H1jet,H1Roma}. The present
MC codes do rely upon the discredited
Regge factorization too heavily , though \cite{Ada}.


\section{Inclusive-exclusive duality in diffractive DIS}

A duality between diffraction into the low mass continuum,
$M^{2} \lsim M_{o}^{2} \sim M_{V}^{2}$,
and exclusive production of vector mesons,
\be
\int_{0}^{M_{o}^{2}} dM^{2}
(d \sigma_{T,L}^{D}/dM^{2})\approx
\sigma_{T,L}(\gamma^{*}\rightarrow V)\,,
\label{eq:7.1}
\ee
is not a conjecture, but a highly nontrivial result derived
from QCD \cite{GNZlong}. Recall the QCD
results \cite{KNNZ94,NNZdipole}
 $\sigma_{T}(\gamma^{*}\rightarrow V) \propto
G^{2}(x,{1\over 4}(Q^{2}+M_{V}^{2}))/(Q^{-2}+M_{V}^{2})^{4}$ and
$\sigma_{L}/\sigma_{T} \approx Q^{2}/M_{V}^{2}$. With our results for
the large-$\beta$ behaviour of $F_{T,L}^{D}(Q^{2},x_{\Pom},\beta)$,
the l.h.s. and r.h.s. of (\ref{eq:7.1}) have a perfectly matching
$Q^{2}$ dependence. It is remarkable how for the $\propto
(1-\beta)^{2}$ decrease of the leading twist $F_{T}^{D}$, the higher
twist $F_{L}^{D}$ which is finite at large-$\beta$, takes over
in the duality integral so that $d\sigma_{L}^{D}/d\sigma_{T}^{D} \approx
Q^{2}/M_{V}^{2}$ for the low-mass continuum! Furthermore, in the
limit of $M^{2} \sim M_{V}^{2}$
we have $q_{T,L}^{2}
\sim {1\over 4}(Q^{2}+M_{V}^{2})$, which matches perfectly the
pQCD scale in the vector meson production \cite{NNZdipole}.
\\

For the {\bf Conclusions} see Section 1.
\\

\end{document}